\documentclass[11pt]{article}
\usepackage{amsmath}

\def\be{\begin{equation}}
\def\ee{\end{equation}}
\def\p{^{\prime}}
\def\a{\alpha}

\def\n{\nabla}
\def\t{\tau}
\def\mb{\mathbf}
\def\v{\nu}
\def\m{\mu}
\def\s{\sigma}
\def\mbx{\mathbf{x}}
\def\o{\omega}

\def\pa{\partial}

\def\ra{\rightarrow}
\def\c{\chi}
\def\O{\Omega}
\title{Massive bosons and the dS/CFT correspondence}
\author{Ois\'{\i}n A. P. Mac Conamhna\thanks{O.A.P.MacConamhna@damtp.cam.ac.uk}
\\DAMTP\\ Centre for Mathematical Sciences\\ University of Cambridge\\ Wilberforce Road, Cambridge CB3 0WA, UK.}

\begin{document}

\maketitle

\begin{abstract}
We compute the boundary two point functions of operators corresponding
to massive spin 1 and spin 2 de Sitter fields, by an extension of the
``S-Matrix'' approach developed for bulk scalars. In each case the
two point functions are of the form required for conformal invariance
of the dual boundary field theory. We emphasise that in the context of
dS/CFT one should consider unitary representations of the Euclidean
conformal group, without reference to analytic continuation of the
boundary theory to Lorentzian signature.

\end{abstract}

\section{Introduction}

The holographic principle is an attractive and fruitful way of
thinking about quantum gravity, which has proven spectacularly
successful in its application to the physics of AdS
spacetimes. Fuelled by this success it is natural to attempt to find a
similar holographic description of spacetimes with positive
cosmological constant. However in spite of
considerable recent activity \cite{hull, witten, strom1, bala, klemm, strom3, cai, strom2, spradlin, abda}, the proposed dS/CFT correspondence
remains very much the poor relation of its AdS counterpart. Numerous
conceptual issues remain to be addressed, chief among which is the
delicate problem of observables in a spacetime with cosmological
horizons. The abscence of a controlled de Sitter solution of string
theory means that we currently have no concrete input on the CFT side
of the correspondence. The precise nature of the correspondence
with the as yet unknown CFT has also remained vague; in particular, there
are difficulties in attempting to define an AdS-like prescription
\cite{spradlin}, \cite{deB}. Furthermore though much work has been
done on bulk scalars, a CFT description of higher spin bulk fields has
remained unexplored.

Here we will address the last issue. Tests of the proposed
correspondence can only be performed with our current level of
knowledge from the bulk side, and this provides the motivation for
seeking a dual description of higher spin bulk fields. In Section 2 we will briefly
review the calculation of CFT two point functions corresponding to
bulk scalars. In sections 3 and 4 we extend the ``S-Matrix'' approach
to massive spin 1 bulk fields in any dimension  and spin 2 bulk fields in $d=4$, and find that in each case
the two point functions obtained are of the form required for
conformal invariance. Section 5 contains a discussion of our results,
the issue of unitarity of the boundary CFT, and  
interesting future directions.

In what follows we will use two coordinatisations of $dS_d$. The
first which we will refer to as global coordinates is given by
\be
ds^2=-d\t^2+\cosh^2\t d\Omega_{d-1}^2
\end{equation}
with $d\O_{d-1}^2=\gamma_{ij}d\o^id\o^j$, $\gamma_{ij}$ the round
metric on $S^{d-1}$. We have set the de Sitter length $l=1$. These
coordinates cover all of $dS_d$. The second set which we
will refer to as planar coordinates is given by
\be
ds^2=-dt^2+e^{-2t}d\mathbf{x}^2
\end{equation}
These cover only half the spacetime, namely the causal past of a
geodesic observer. We also define the function 
\be
z(x,x\p)=\cos^2\frac{\m(x,x\p)}{2}
\end{equation}
where $\m(x,x\p)$ is the geodesic distance between two points.

\section{Bulk Scalars}
First we consider bulk scalar fields. We will for simplicity take
$d=3$. This has been extensively studied in \cite{strom1}, \cite{strom2},
\cite{spradlin}, and we will review their results. The equation of motion is
\be
(\n^2-m^2)\phi=0
\end{equation} 
The limiting behaviour of $\phi$ in global coordinates as
$\t\ra-\infty$  is
\be
\lim_{\t\to
-\infty}\phi(\t,\Omega)=\phi_{+}^{\mbox{in}}(\Omega)e^{h_+\t}+\phi_{-}^{\mbox{in}}(\Omega)e^{h_-\t}
\end{equation}
where $h_{\pm}=1\pm \v$, $\v=\sqrt{1-m^2}$. The Wightman function
$G(x,x\p)$ is vacuum dependent and is a linear combination
of the hypergeometric functions
\be
F\Big(h_+,h_-;\frac{3}{2};z\Big),\:F\Big(h_+,h_-;\frac{3}{2};1-z\Big)
\end{equation}
We consider the amplitude
\be\label{beld}
(\phi,\phi)_{--}=\lim_{\t\to-\infty}\int\int d^2\Omega
d^2\Omega\p\Big[\sqrt{-g}\sqrt{-g\p}\Big(\phi(x)\overleftrightarrow{\partial}_{\t}G(x,x\p)\overleftrightarrow{\partial}_{\t\p}\phi(x\p)\Big)\Big]_{\t=\t\p}
\end{equation}
and find on choosing a particular vacuum 
\be\label{gowk}
(\phi,\phi)_{--}=\int\int d^2\Omega d^2\Omega\p
\Big(b_-\phi_{+}^{\mbox{in}}(\Omega)\Delta_-(\Omega,\Omega\p)\phi_+^{\mbox{in}}(\Omega\p)+b_+\phi^{\mbox{in}}_-(\O)\Delta_+(\O,\O\p)\phi^{\mbox{in}}_-(\O\p)\Big)
\end{equation}
with $b_{\pm}$ vacuum dependent constants (which vanish in the ``in'' vacuum). The kernels
$\Delta_{\pm}(\O,\O\p)$ are the correlation functions of CFT operators
of conformal dimension $h_{\pm}$ on an $S^2$. The constants  may be
absorbed by a field redefinition for two point functions, though three
and higher point functions depend nontrivially on the choice of vacuum
\cite{spradlin}. Similarly one may evaluate $(\phi,\phi)_{-+}$,
$(\phi,\phi)_{+-}$ and $(\phi,\phi)_{++}$. In particular with one point
on $\mathcal{I}^-$ and one point on $\mathcal{I}^+$, we define
$\phi^{\mbox{out}}$ with an antipodal inversion relative to
$\phi^{\mbox{in}}$:
\be
\lim_{\t\to\infty}\phi(\t,\O_{A})=\phi^{\mbox{out}}_+(\O)e^{-h_+\t}+\phi^{\mbox{out}}_-(\O)e^{-h_-\t}
\end{equation}
and then
\be\label{bug}
(\phi,\phi)_{+-}=\int\int d^2\Omega d^2\Omega\p
\Big(\hat{b}_-\phi_+^{\mbox{out}}(\Omega)\Delta_-(\Omega,\Omega\p)\phi_+^{\mbox{in}}(\Omega\p)+(-\leftrightarrow+)\Big)
\end{equation}
When evaluating CFT correlators with one point on each boundary, there
is a wider choice of bulk Green functions that may be used in
(\ref{beld}), as the bulk Schwinger function is then nonzero. In
\cite{strom2} the asymptotic behaviour of the bulk Wightman function
with one point on each boundary is considered, while in the
S-Matrix interpretation of \cite{spradlin} it is the Feynman
propagator that is used. The dependence of CFT correlators on the bulk
vacuum is encoded in the asymptotic behaviour of the bulk Hadamard
function, as the Schwinger function is vacuum independent.

\section{Massive Spin 1 Bulk Fields} 
We will apply a similar analysis to massive spin 1 bulk
fields. The two point
functions of the boundary operators may be obtained up to a constant
factor as the kernels of
the bilinear forms obtained from the amplitude (in planar coordinates)
\begin{equation}\label{amp}
(A,\:A)=\lim_{t\to -\infty}\int\int d^{d-1}x d^{d-1}x^{\prime} \Big[\sqrt{-g}\sqrt{-g^{\prime}}g^{\m\v}g^{\m^{\prime} \v^{\prime}}\Big(A_{\m}(t,\mb{x})\overleftrightarrow{\nabla}_tG_{\v\v^{\prime}}(t, \mb{x}, t^{\prime}, \mb{x}^{\prime})\overleftrightarrow{\nabla}_{t^{\prime}}A_{\m^{\prime}}(t^{\prime},\mb{x}^{\prime})\Big)\Big]_{t=t^{\prime}}
\end{equation}
where $G_{\m \v^{\prime}}(t, \mb{x,}t^{\prime},\mb{x}^{\prime})$
is the Wightman function of $A_{\m}(t,\mb{x})$
\begin{equation}
G_{\m \v^{\p}}(t, \mb{x},
t^{\prime},\mb{x}^{\prime})=\langle0|A_{\m}(t,\mb{x})A_{\v^{\prime}}(t^{\prime},\mb{x}^{\prime})|0\rangle
\end{equation}
with $|0\rangle$ a de Sitter invariant vacuum state. The equation of motion for $A_{\v}$ is 
\begin{equation}\label{eom}
(\nabla^{2}-(d-1)-m^{2})A_{\v}=0
\end{equation}
subject to the constraint
\begin{equation}
\nabla_{\mu}A^{\mu}=0
\end{equation}
We solve these equations asymptotically. The constraint equation yields
\begin{equation}\label{con}
\partial_{t}A_{t}-(d-1)A_{t}=e^{2t}D_{i}A^{i}
\end{equation} 
(\ref{eom}) yields
\begin{eqnarray}\label{A0}
\partial_{t}^{2}A_{t}-e^{2t}D^{2}A_t-2e^{2t}
D_{i}A^{i}-(d-1)\partial_{t}A_{t}+m^{2}A_t&=&0\\
\label{scal}\partial_{t}^{2}A_{i}-e^{2t}D^{2}A_{i}
-(d-3)\partial_{t}A_{i}-2
D_{i}A_{t}+m^{2}A_{i}&=&0
\end{eqnarray}  
Combining (\ref{con}) and (\ref{A0}) we see that as
$t\rightarrow -\infty$ 
\begin{equation}\label{assa0}
A_{t}(t,\mb{x})\rightarrow
e^{(r_{-}+2)t}A_{-t}(\mb{x})+e^{(r_{+}+2)t}A_{+t}(\mb{x})
\end{equation}
where
\begin{eqnarray}
r_{\pm}&=&\frac{d-3\pm \rho}{2}\nonumber\\\rho&=&\sqrt{(d-3)^2-4m^2}
\end{eqnarray}
Then from (\ref{scal}) 
\begin{equation}\label{assa}
A_{i}(t ,\mb{x})\rightarrow
e^{r_{-}t}A_{-i}(\mb{x})+e^{r_{+}t}A_{+i}(\mb{x})
\end{equation}
with all corrections suppressed by at least two powers of
$e^{t}$. The Wightman function
$G_{\m\v^{\prime}}(x, x^{\prime})$ satisfies
\begin{equation}
(\nabla^{2}-(d-1)-m^{2})G_{\m\v^{\prime}}(x, x^{\prime})=0
\end{equation}
subject to
\begin{equation}
\nabla_{\mu}G^{\mu}_{\v^{\prime}}(x, x^{\prime})=0
\end{equation}
When a maximally symmetric vacuum state is chosen $G_{\m\v^{\prime}}$ is a maximally
symmetric bivector. This means that it may be expressed as
the sum of products of a set of preferred geometric objects
\cite{allen}: the geodesic distance $\mu(x,x\p)$ (equivalently $z(x,x\p)$); the unit tangents to
the geodesic at $x$ ($\n_{\a}\mu(x,x\p)\equiv n_{\a}(x,x\p)$) and at $x\p$
($\n_{\v\p}\mu(x,x\p)\equiv n_{\v\p}(x,x\p)$); and the
parallel propagator along the geodesic
$g^{\m}_{\;\v\p}(x,x\p)$. $g_{\m\v\p}$ satisfies
\begin{eqnarray}
g_{\!\m}^{\;\v\p}(x,x\p)g_{\v\p\sigma}(x\p,x)&=&g_{\m\sigma}(x)\\
g_{\!\m\p}^{\;\v}(x\p,x)g_{\v\sigma\p}(x,x\p)&=&g_{\m\p\sigma\p}(x\p)
\end{eqnarray}
with $g_{\m\v}$, $g_{\m\p\v\p}$ the metric at $x,\:x\p$. $g_{\s\v\p}$
may be expressed in terms of $z$, $n_{\sigma}$ and $\:n_{\v\p}$ as
\be
g_{\sigma\v\p}=-2\sqrt{z(1-z)}\n_{\sigma}n_{\v\p}-n_{\sigma}n_{\v\p}
\end{equation}
We will now choose the Euclidean vacuum state. This is the unique
de Sitter invariant vacuum state where the Wightman function is
singular for coincident points but nonsingular for antipodal points on
the de Sitter hyperboloid. It would be interesting
to examine the r\^{o}le of other vacua, but such issues are not explored
here. Then the Wightman function is given by \cite{allen}
\begin{eqnarray}\label{ai}
G_{\sigma\v^{\prime}}(x,x^{\prime})&=&f(z)g_{\s\v^{\prime}}+g(z)n_{\s}n_{\v^{\prime}}\\f(z)&=&\Big(-\frac{2}{d-1}z(1-z)\frac{d}{dz}+2z-1\Big)u(z)\nonumber\\g(z)&=&f(z)-u(z)\nonumber\\u(z)&=&qF\Big(r_++2,r_-+2;\:\frac{d+2}{2};z\Big)\nonumber\\q&=&\frac{(1-d)\Gamma(r_++2)\Gamma(r_-+2)}{2^{d+1}\pi^{d/2}\Gamma(\frac{d+2}{2})m^2}
\end{eqnarray}
The constant $q$ is chosen such that the short distance singularity
matches that of flat space. Expressing in terms of $P=2z-1$ and retaining
only the terms which contribute to (\ref{amp}) as
$t\rightarrow-\infty$ we find
\begin{equation}
G_{\s\v^{\prime}}(x,x^{\prime})=\Big[\Big(\frac{P^2}{d-1}\frac{d}{dP}+P\Big)u(P)\Big]\Big[-\frac{\nabla_{\s}P\n_{\v^{\prime}}P}{P}+\nabla_{\s}\nabla_{\v^{\prime}}P\Big]
\end{equation}
As $t\rightarrow-\infty$, $P(x,x^{\prime})\rightarrow
-\frac{e^{-(t+t^{\prime})}}{2}|\mathbf{x-x^{\prime}}|^2
$
and hence 
\begin{eqnarray}
-\frac{\nabla_iP\nabla_{j^{\prime}}P}{P}+\nabla_i\nabla_{j^{\prime}}P&\rightarrow&
e^{-(t+t^{\prime})}\Big(\delta_{ij^{\prime}}-\frac{2(x-x^{\prime})_i(x-x^{\prime})_{j^{\prime}}}{|\mathbf{x-x^{\prime}}|^2}\Big)\nonumber\\&\equiv&e^{-(t+t^{\prime})}I_{ij^{\prime}}(\mbx-\mbx\p)
\end{eqnarray}
Furthermore the coefficient of the leading $e^{-(t+t\p)}$ term in
$-P^{-1}\n_tP\n_{\v\p}P+\n_t\n_{\v\p}P$ is zero, and hence
$G_{t\v\p}(x,x\p)$ is suppressed by at least two powers of $e^t$,
$e^{t\p}$ relative to $G_{ij\p}(x,x\p)$. The asymptotic behaviour of $G_{ij\p}(x,x\p)$ is
determined by the transformation properties of the hypergeometric
function (see appendix). We find that
\be\label{assg}
G_{ij\p}(x,x\p)\rightarrow
\Big(\frac{a_-e^{r_-(t+t\p)}}{|\mb{x}-\mb{x}\p|^{2(r_-+1)}}+\frac{a_+e^{r_+(t+t\p)}}{|\mb{x}-\mb{x}\p|^{2(r_++1)}}\Big)I_{ij\p}
\end{equation} 
$a_{\pm}$ constants. Then inserting (\ref{assa0}), (\ref{assa}) and
(\ref{assg}) into (\ref{amp}) and ignoring contact terms, we find
\be\label{yik}
(A,\:A)=\int\int
d^{d-1}xd^{d-1}x\p\Big(\frac{\hat{a}_+A^{i}_-(\mbx)A^{j\p}_-(\mbx\p)}{|\mbx-\mbx\p|^{2(r_++1)}}+\frac{\hat{a}_-A^{i}_+(\mbx)A^{j\p}_+(\mbx\p)}{|\mbx-\mbx\p|^{2(r_-+1)}}\Big)I_{ij\p}
\end{equation}
where $A_{\pm}^{ i}=\delta^{ij}A_{\pm j}$ and 
\be
\hat{a}_{\pm}=-\frac{q\rho^2(r_{\pm}-1)\Gamma(\mp\rho)\Gamma((d+2)/2)4^{r_{\pm}+2}}{6\Gamma(r_{\mp}+2)\Gamma((1\mp\rho)/2)}
\end{equation}
Hence we see that $A_{\pm}^i$ correspond to a pair of CFT operators
$V^i_{\pm}$ whose two point functions are up to a constant the kernels
of (\ref{yik}):
\be\label{gaz}
<V_{\pm}^i(\mbx)V_{\pm}^{j\p}(\mbx\p)>
=\frac{I^{ij\p}(\mbx-\mbx\p)}{|\mbx-\mbx\p|^{2(r_{\pm}+1)}}
\end{equation}
which is of the form required for conformal invariance\footnote{We
note that this derivation is valid in any dimension if the mass of the
bulk field is such that $\rho$ is purely imaginary or $\rho$ is real
and $\rho<2$. If $\rho>2$ (as may occur if $d\ge6$) then corrections
to the leading asymptotic behaviour of the field could in principle
contibute to (\ref{amp}), and a more detailed analysis is
required. These corrections are nevertheless not expected to change
the result. The cases when $\rho$ is an integer require a case by case
treatment and are explicitly excluded from the discussion.}. The conformal
dimensions  are 
\be
\frac{1}{2}(d-1\pm\sqrt{(d-3)^2-4m^2})
\end{equation} In global coordinates it may readily be verified that the asymptotic
behaviour of $A_{i}$ is given by
\begin{equation}
\lim_{\t\to-\infty}A_{i}(\t,\mb{\Omega})=e^{
r_-\t}A_{-i}^{\mbox{in}}(\mb{\Omega})+e^{r_+\t}A^{\mbox{in}}_{+i}(\mb{\Omega})
\end{equation} 
with $A_{\t}(\t,\O)$ suppressed by two powers of $e^{\t}$ as
before. As $\t$, $\t\p\rightarrow-\infty$ we find that
$P(x,x\p)\rightarrow-\frac{e^{-(\t+\t\p)}}{2}\Big(\frac{1-\cos\theta}{2}\Big)$,
where $\theta(\mb{\O},\mb{\O}\p)$ is the geodesic distance on
the $S^{d-1}$. Also we see that
\begin{eqnarray}
-\frac{\n_iP\n_{j\p}P}{P}+\n_i\n_{j\p}P&\rightarrow&
\frac{e^{-(\t+\t\p)}}{4}(\n_i\theta
\n_{j\p}\theta-\sin\theta\n_i\n_{j\p}\theta)\nonumber\\&\equiv& \frac{e^{-(\t+\t\p)}}{4}\mathcal{I}_{ij\p}
\end{eqnarray}
$\mathcal{I}_{ij\p}$ generalises the flat space inversion tensor $I_{ij\p}$ to
the unit sphere \cite{hugh1}. We define $\s=\frac{1-\cos\theta}{2}$,
the generalisation of the flat space $|\mbx-\mbx\p|^2$. Then
evaluating the analogue of (\ref{amp}) in global coordinates as $\t$,
$\t\p\rightarrow-\infty$ we find
\be
(A,A)_{--}=\frac{1}{2^{d+1}}\int\int d^{d-1}\O d^{d-1}\O\p\Big(\frac{\hat{a}_-A^{\mbox{in}}_{+i}(\mb\O)A^{\mbox{in}}_{+j\p}(\mb\O\p)}{\s^{(r_-+1)}}+\frac{\hat{a}_+A^{\mbox{in}}_{-i}(\mb\O)A^{\mbox{in}}_{-j\p}(\mb\O\p)}{\s^{(r_++1)}}\Big)\mathcal{I}^{ij\p}
\end{equation}
where $\mathcal{I}^i_{k\p}=\gamma^{ij}\mathcal{I}_{jk\p}$. The
kernels are proportional to the two point functions of the CFT
operators and are just (\ref{gaz}) conformally rescaled to the sphere. Similarly one may evaluate $(A,A)_{++}$, $(A,A)_{-+}$ and
$(A,A)_{+-}$, in particular finding the familiar antipodal
inversion when one point is on $\mathcal{I}^+$ and
one on $\mathcal{I}^-$.

\section{Massive Spin 2 Fields}
We follow a similar procedure for massive spin 2 fields, which were discussed in \cite{deser}. In four dimensions the equation
of motion is
\be\label{y}
(\n^2-(m^2+2))\c_{\m\v}=0
\end{equation}
subject to the constraints
\be\label{z}
\n_{\m}\c^{\m}_{\v}=0,\:g^{\m\v}\c_{\m\v}=0
\end{equation}
and $\c_{\m\v}$ symmetric. Bulk unitarity requires $m^2\ge 2$, with
$m^2=2$ giving partial masslessness (in general dimension partial
masslessness occurs at $m^2=d-2$ \cite{buch}). We assume $m^2>2$. In global coordinates, (\ref{y}) and
(\ref{z}) read
\begin{eqnarray}
&&\partial_{\t}^2\c_{\t\t}-\frac{D^2\c_{\t\t}}{\cosh^2\t}+7\tanh\t\partial_{\t}\c_{\t\t}+(8\tanh^2\t+2+m^2)\c_{\t\t}=0\nonumber\\&&\partial_{\t}^2\c_{\t
i}-\frac{D^2\c_{\t i}}{\cosh^2\t}+\tanh\t(3\partial_{\t}\c_{\t
i}+2\partial_i\c_{\t\t})+\Big(\frac{1}{\cosh^2\t}+m^2\Big)\c_{\t
i}=0\nonumber\\&&\pa^2_{\t}\c_{ij}-\frac{D^2\c_{ij}}{\cosh^2\t}-\tanh\t\pa_{\t}\c_{ij}+4\tanh\t
D_{(i}\c_{j)\t}-(2\tanh^2\t-m^2)\c_{ij}-2\sinh^2\t\gamma_{ij}\c_{\t\t}=0\nonumber\\&&\pa_{\t}\c_{\t\t}-\frac{\gamma^{ij}D_i\c_{j\t}}{\cosh^2\t}+3\tanh\t\c_{\t\t}+\frac{\tanh\t\gamma^{ij}\c_{ij}}{\cosh^2\t}=0\nonumber\\&&\pa_{\t}\c_{\t
i}-\frac{\gamma^{jk}D_j\c_{ki}}{\cosh^2\t}+3\tanh\t\c_{\t
i}=0\nonumber\\&&\c_{\t\t}-\frac{\gamma^{ij}\c_{ij}}{\cosh^2\t}=0
\end{eqnarray}
and these give the asymptotic behaviour as $\t\ra-\infty$
\begin{eqnarray}
\c_{\t\t}(\t,\O)&\ra&
e^{(q_-+2)\t}\c_{\t\t}^{-}(\O)+e^{(q_++2)\t}\c_{\t\t}^+(\O)+O(e^{(q_{\pm}+4)\t})\nonumber\\\c_{\t
i}(\t,\O)&\ra& e^{q_-\t}\c^-_{\t
i}(\O)+e^{q_+\t}\c^+_{\t
i}(\O)+O(e^{(q_{\pm}+2)\t})\nonumber\\\c_{ij}(\t,\O)&\ra&
e^{(q_--2)\t}\c^-_{ij}(\O)+e^{(q_+-2)\t}\c^+_{ij}(\O)+O(e^{q_{\pm}\t})\nonumber\\\label{poo}\gamma^{ij}\c_{ij}^-&=&\gamma^{ij}\c_{ij}^+=0
\end{eqnarray}
where
\be
q_{\pm}=\frac{3\pm\sqrt{9-4m^2}}{2}
\end{equation}
Following \cite{grav} it is straightforward to solve for the Wightman
function $G_{\m\v\a\p\s\p}(x,x\p)$ in terms of $z$. Since it is
traceless on primed and unprimed indices,
$G_{\m\v\a\p\s\p}$ may be expressed in terms of a traceless basis of
bitensors 
\be
\mb{G}(x,x\p)=X(\m)\mb{T}^1+Y(\m)\mb{T}^2+Z(\m)\mb{T}^3
\end{equation}
where
\begin{eqnarray}
T^1_{\m\v\a\p\s\p}&=&\frac{1}{16}g_{\m\v}g_{\a\p\s\p}+n_{\m}n_{\v}n_{\a\p}n_{\s\p}-\frac{1}{4}(g_{\m\v}n_{\a\p}n_{\s\p}+n_{\m}n_{\v}g_{\a\p\s\p})\nonumber\\T^2_{\m\v\a\p\s\p}&=&(g_{\m\a\p}g_{\v\s\p}+g_{\m\s\p}g_{\v\a\p})-\frac{1}{2}g_{\m\v}g_{\a\p\s\p}\nonumber\\T^3_{\m\v\a\p\s\p}&=&4n_{(\m}g_{\v)(\a\p}n_{\s\p)}+4n_{\m}n_{\v}n_{\a\p}n_{\s\p}
\end{eqnarray}
Defining the functions 
\begin{eqnarray}
W&=&X+\frac{8}{3}Y\nonumber\\U&=&Y-Z,
\end{eqnarray}
the equations of motion and constraints yield, in terms of the variable
$z$,
\be
\Big(z(1-z)\frac{d^2}{dz^2}+[c-(1+a+b)z]\frac{d}{dz}-ab\Big)W=0
\end{equation}
and
\begin{eqnarray}
U&=&\frac{3}{16}\Big(z(z-1)\frac{dW}{dz}+2(2z-1)W\Big)\nonumber\\Y&=&\frac{3}{40}\Big(8z(z-1)\frac{dU}{dz}+16(2z-1)U-W\Big)
\end{eqnarray}
where $a=q_-+2$, $b=q_++2$ and $c=4$, so in the Euclidean vacuum, $W=F(q_-+2,q_++2;4;z)$ (up to a constant)
and  $X,\:Y$ and $Z$ are completely determined.
It is convenient for our purposes to change basis, and express the
Wightman function as 
\be
\mb{G}(x,x\p)=\frac{1}{16}(X-8Y)\mb{Q}_1+(X+2Y)\mb{Q}_2+Y\mb{Q}_3-\frac{1}{4}X\mb{Q}_4+U\mb{Q}_5
\end{equation}
The new basis bitensors are
\begin{eqnarray}
Q^1_{\m\v\a\p\s\p}&=&g_{\m\v}g_{\a\p\s\p}\nonumber\\Q^2_{\m\v\a\p\s\p}&=&n_{\m}n_{\v}n_{\a\p}n_{\s\p}\nonumber\\Q^3_{\m\v\a\p\s\p}&=&4z(1-z)(\n_{\m}n_{\a\p}\n_{\v}n_{\s\p}+\n_{\m}n_{\s\p}\n_{\v}n_{\a\p})\nonumber\\Q^4_{\m\v\a\p\s\p}&=&g_{\m\v}n_{\a\p}n_{\s\p}+n_{\m}n_{\v}g_{\a\p\s\p}\nonumber\\Q^5_{\m\v\a\p\s\p}&=&8\sqrt{z(1-z)}n_{(\m}(\n_{\v)}n_{(\a\p})n_{\s\p)}
\end{eqnarray}
The leading asymptotic contribution to the Wightman function comes
from the $ijk\p l\p$ component of the $\mb{Q}_1$, $\mb{Q}_3$ terms. We find
\begin{eqnarray}\label{16}
\Big(\frac{1}{16}(X-8Y)Q_1+YQ_3\Big)_{ijk\p
l\p}&\ra&\Big(\frac{c_-e^{(q_--2)(\t+\t\p)}}{\s^{q_-}}+\frac{c_+e^{(q_+-2)(\t+\t\p)}}{\s^{q_+}}\Big)\Big(\mathcal{I}_{ik\p}\mathcal{I}_{jl\p}+\mathcal{I}_{il\p}\mathcal{I}_{jk\p}-\frac{2}{3}\gamma_{ij}\gamma_{k\p
l\p}\Big)\nonumber\\&\equiv& c_-e^{(q_--2)(\t+\t\p)}H^-_{ijk\p
l\p}+(-\leftrightarrow +)
\end{eqnarray}
with all corrections and other components suppressed by at least two
powers, and $c_{\pm}$ constants. (\ref{16}) is traceless at the boundary
$(\gamma^{ij}\mathcal{I}_{ik\p}\mathcal{I}_{j l\p}=\gamma_{k\p l\p})$
since $\c_{ij}^{\pm}$ are traceless there (\ref{poo}). The CFT two
point functions are given by the kernels of
\be
(\c,\c)_{--}=\lim_{\t\to -\infty}\int\int d^3\O
d^3\O\p\Big[\sqrt{g(x)g(x\p)}g^{\a\m}g^{\s\v}\Big(\c_{\a\s}(x)\overleftrightarrow{\nabla}_{\t}G_{\m\v\a\p\s\p}(x,x\p)\overleftrightarrow{\n}_{\t\p}\c_{\m\p\v\p}(x\p)\Big)g^{\a\p\m\p}g^{\s\p\v\p}\Big]_{\t=\t\p}
\end{equation}
A careful counting of powers reveals that (\ref{16}) provides the only
contribution:
\be
(\c,\c)_{--}=\int\int d^3\O
d^3\O\p\Big(\hat{c}_-\chi^{+ij}(\O)H^-_{ijk\p l\p}(\O,\O\p)\chi^{+k\p
l\p}(\O\p)+(-\leftrightarrow
+)\Big)
\end{equation}
Thus the CFT correlators are determined up to a constant as
\be
<T^{\pm}_{ij}(\O)T^{\pm}_{k\p
l\p}(\O\p)>=H^{\pm}_{ijk\p l\p}(\O,\O\p)
\end{equation}
which are the two point functions required by conformal invariance for
traceless symmetric tensor fields of conformal dimension
$\frac{1}{2}(3\pm\sqrt{9-4m^2})$ on an $S^3$ (the flat space form is
given in eg \cite{palchik}). The
calculation may be trivially modified to planar coordinates, or to
involve one point on each boundary. In general we expect
conformal dimensions $\frac{1}{2}(d-1\pm\sqrt{(d-1)^2-4m^2})$.

\section{Discussion}
In conclusion we have tested dS/CFT for higher spin massive bulk
bosons. The symmetric traceless tensor representations of $SO_0(d,\:1)$
(the identity component of the Euclidean conformal group in $d-1$
dimensions) are labelled by $[s,\:\Delta]$, where $s$ is the tensor
rank and $\Delta$ the conformal dimension. It is well known
\cite{shite}, \cite{buck} that some
of these representations may  be continued to unitary representations
of the universal covering of $SO_0(d-1,\:2)$. This is possible if
$\Delta$ is real and 
\begin{eqnarray}
\Delta&\ge&\frac{d-3}{2},\:\:\:s=0\nonumber\\\Delta&\ge& d-3+s,\:\:\:s\ge
1
\end{eqnarray}
That these conditions are violated in the context of dS/CFT by all but a small range of scalar
representations has led to the suggestion
that the putative CFT dual is nonunitary. However as the boundary CFT
is a priori Euclidean, it is more natural to demand the unitary
realisation of $SO_0(d,\:1)$ symmetry with no reference made to
analytic continuation to Lorentzian signature. Apart from certain
exceptional, topologically reducible representations, the unitary
symmetric traceless tensor representations of $SO_0(d,\:1)$ (discussed in detail in \cite{buck})
are irreducible and unique up to equivalence. They are given by the
following. 

The principal series:
\be
s=0,1,2...\;;\:\:\:\:\:\Delta=\frac{d-1+i\s}{2},\:\:\:\:\:\s\:\:\mbox{real}
\end{equation}

The complementary series:
\begin{eqnarray}
s=0;\:\:\:0<\Delta<d-1\:\:\:\:\:(d-1\ge2)\\\label{eeek}s\ge 1\;;\:\:\:1<\Delta<d-2\:\:\:\:\:(d-1\ge3)
\end{eqnarray}
In the bulk the principal and complementary series of representations
of $SO_0(d,\:1)$ are distinguished by the value of the
mass. Taking for example the scalar representations, in $d$ bulk
dimensions we have $m^2>(d-1)^2/4$ for the principal series and
$0<m^2<(d-1)^2/4$ for the complementary series. Principal (complementary) series representations in the bulk
correspond to principal (complementary) series representations on the
boundary. Similar considerations hold for spin 1 and spin 2 (
\cite{gaz1}, \cite{gaz2} and references therein). In particular  we note that the bulk unitarity bound on the mass of
the spin 2 field has a holographic reflection in (\ref{eeek}). We
conclude that the group theoretic content of dS/CFT is that bulk and
boundary representations of $SO_0(d,\:1)$ are unitarily equivalent. 

It would be interesting to extend the calculations of CFT correlators
to the exceptional representations mentioned above. These should correspond to
bulk photons, gravitons
and partially massless spin 2 fields, among others. 
 However there are a number of difficulties which
need to be overcome: there are subtleties in a de
Sitter invariant treatment of massless particles and also a problem
with infrared divergences of the Green functions \cite{tolley},
\cite{mottola}. Another problem is to obtain a better understanding of
field theories based on unitary representations of the Euclidean
conformal group (this has been examined in two dimensions in
\cite{deB}, where novel hermiticity conditions were introduced). It
would also be of great interest to find
a more precise definition, going beyond the current S-Matrix
formalism, of what is meant by dS/CFT.

\section{Acknowledgements}
I am grateful to Francis Dolan, Hugh Osborn and in particular Stephen
Hawking. This work was supported by the National University of
Ireland, EPSRC, and a Freyers Studentship.

\section{Appendix}
Here we relate properties of the hypergeometric function used in the
text.
\begin{eqnarray}
F(a,b;c;z)=\frac{\Gamma(c)\Gamma(b-a)}{\Gamma(b)\Gamma(c-a)}(-z)^{-a}F(a,1-c+a;1-b+a;z^{-1})\nonumber\\+\frac{\Gamma(c)\Gamma(a-b)}{\Gamma(a)\Gamma(c-b)}(-z)^{-b}F(b,1-c+b;1-a+b;z^{-1})
\end{eqnarray}
\be
\lim_{z\to\infty}F(a,b;c;z^{-1})=1+O(z^{-1})
\end{equation}
\begin{eqnarray}
\frac{d^n}{dz^n}F(a,b;c;z)&=&\frac{(a)_n(b)_n}{(c)_n}F(a+n,b+n;c+n;z)\\(a)_n&=&\frac{\Gamma(a+n)}{\Gamma(a)}
\end{eqnarray}


\begin{thebibliography}{99}
\bibitem{hull} C. Hull, JHEP 9807 021, 1998
\bibitem{witten} E. Witten, hep-th/0106109
\bibitem{strom1} A. Strominger, JHEP 0110 034, 2001
\bibitem{bala} V. Balasubramanian, J. de Boer and D. Minic,
Phys. Rev. D65 (2002) 123508
\bibitem{klemm} D. Klemm, Nucl. Phys. B625 (2002) 295 
\bibitem{strom3} A. Strominger, JHEP 0111 049, 2001
\bibitem{cai} R-G. Cai, Phys. Lett. B525 (2002) 331
\bibitem{strom2} R.Bousso, A. Maloney and A. Strominger,
Phys. Rev. D65 (2002) 104039
\bibitem{spradlin} M. Spradlin and A. Volovich, Phys. Rev. D 65 (2002)
104037

\bibitem{abda} E. Abdalla, B. Wang, A. Lima-Santos, W. G. Qiu,
Phys. Lett. B538 (2002) 435
\bibitem{deB} V. Balasubramanian, J. de Boer and D. Minic, hep-th/0207245
\bibitem{allen} B. Allen and T. Jacobsen, Comm. Maths. Phys. 103 (1986)
669-692
\bibitem{hugh1} H. Osborn and G. M. Shore, Nucl. Phys. B571 (2000) 287
\bibitem{deser} S. Deser and A. Waldron, Nucl. Phys. B607 (2001) 577
\bibitem{buch} I. L. Buchbinder, D. M. Gitman and V. D. Pershin,
Phys. Lett. B492 (2000) 161
\bibitem{grav} B. Allen and M. Turyn, Nucl. Phys. B292 (1987) 813
\bibitem{palchik} E. S. Fradkin and M. Ya. Palchik, hep-th/9712045
\bibitem{shite} E. S. Fradkin and M. Ya. Palchik, Conformal Quantum Field
Theory in D-dimensions, Kluwer Academic Publishers,
1996
\bibitem{buck} V. K. Dobrev, G. Mack, V. B. Petkova, S. G. Petrova and
I. T. Todorov, Lecture notes in Physics, v.63, Springer-Verlag, 1977
\bibitem{gaz1} J-P. Gazeau, J. Renaud and M. Takook,
Class. Quant. Grav. 17 (2000) 1415
\bibitem{gaz2} J-P. Gazeau, M. V. Takook, J. Math. Phys. 41 (2000) 5920
\bibitem{tolley} A. J. Tolley and N. Turok, hep-th/0108119
\bibitem{mottola} I. Antoniadis and E. Mottola, J. Math. Phys. 32 (4)
 (1991) 1037
\end{thebibliography}
\end{document}